\documentclass[aps,prb,10pt,twocolumn]{revtex4-1}
\usepackage{bm}
\usepackage{amsmath}
\usepackage{amsfonts}
\usepackage{amssymb}
\usepackage{float}
\usepackage[final]{graphicx}
\usepackage[caption = false]{subfig}
\usepackage{dsfont}
\usepackage[colorlinks=true,linktocpage=true,breaklinks=true]{hyperref}
\begin{document}
\title{Semi-realistic tight-binding model for Dzyaloshinskii-Moriya interaction}
\author{Ahmed Hajr$^{1,2}$}
\author{Abdulkarim Hariri$^1$}
\author{Guilhem Manchon$^1$}
\author{Sumit Ghosh$^1$}
\author{Aur\'elien Manchon$^{1,3,4}$}
\email{manchon@cinam.univ-mrs.fr}
\affiliation{$^1$King Abdullah University of Science and Technology (KAUST), Physical Science and Engineering Division (PSE), Thuwal 23955-6900, Saudi Arabia.}
\affiliation{$^2$King Fahd University of Petroleum and Minerals (KFUPM), Physics Department, Az Zahran, Saudi Arabia.}
\affiliation{$^3$King Abdullah University of Science and Technology (KAUST), Computer, Electrical and Mathematical Science and Engineering Division (CEMSE), Thuwal 23955-6900, Saudi Arabia.}
\affiliation{$^4$Aix-Marseille Univ, CNRS, CINaM, Marseille, France}
%%%%%%%%%%%%%%%%%%%%%

\begin{abstract}
In this work, we discuss the nature of Dzyaloshinskii-Moriya interaction (DMI) in transition metal heterostructures. We first derive the expression of DMI in the small spatial gradient limit using Keldysh formalism. This derivation provides us with a Green's function formula that is well adapted to tight-binding Hamiltonians. With this tool, we first uncover the role of orbital mixing: using both a toy model and a realistic multi-orbital Hamiltonian representing transition metal heterostructures, we show that symmetry breaking enables the onset of interfacial orbital momentum that is at the origin of the DMI. We then investigate the contribution of the different layers to the DMI and reveal that it can expand over several nonmagnetic metal layers depending on the Fermi energy, thereby revealing the complex orbital texture of the band structure. Finally, we examine the thickness dependence of DMI on both ferromagnetic and nonmagnetic metal thicknesses and we find that whereas the former remains very weak, the latter can be substantial.
\end{abstract}
%\pacs{72.10.-d, 72.15.Jf, 73.23.-b} pll
%\keywords{DMI, Pt/Co interface, SOC}
\maketitle

%%%%%%%%%%%%%%%%%%%%%

\section{Introduction}

Magnetic textures presenting a well-defined chirality are of major interest due to their potential applications in data storage \cite{Fert2013}, brain-inspired architectures \cite{Li2017,Song2019,Zazvorka2019}, and reservoir computing \cite{Prychynenko2018}. Homochiral spin spirals \cite{Ferriani2008,Meckler2009,Menzel2012}, quasi-one dimensional N\'eel walls \cite{Chen2013d,Chen2013c,Tetienne2015}, magnetic skyrmions \cite{Yu2010,Muhlbauer2009,Heinze2011,Romming2013,Chen2015b,Jiang2015,Woo2016,Moreau-Luchaire2016,Boulle2016} in perpendicularly magnetized systems, but also merons in planar magnetic heterostructures \cite{Yu2018,Gobel2019} are currently the object of intense theoretical and experimental investigations as they display high current-velocity characteristics \cite{Yang2015a,Caretta2018}. The key mechanism underlying these magnetic entities is the Dzyaloshinskii-Moriya interaction \cite{Dzyaloshinskii1957,Moriya1960} (DMI), an antisymmetric magnetic exchange that forces neighboring magnetic moments to align perpendicular to each other.\par

In the atomistic limit, where the magnetic moments are localized and well defined, the Dzyaloshinskii-Moriya (DM) energy reads
\begin{eqnarray}\label{eq:dmiat}
E_{\rm DM}=\sum_{ij}{\cal D}_{ij}\cdot({\bf S}_i\times{\bf S}_j),
\end{eqnarray}
where ${\bf S}_{i}$ is the direction of the magnetic moment at site $i$, ${\cal D}_{ij}$ is the DM vector and the sum runs over all the pairs $i,j$ of the system. In this general definition, DMI is not limited to nearest neighbors and from the symmetry viewpoint, ${\cal D}_{ij}$ is determined by Moriya's rules \cite{Moriya1960}. In the micromagnetic limit, where the magnetic order is represented by a continuous vector field ${\bf m}$ with smooth spatial variation, DMI is rewritten
\begin{eqnarray}\label{eq:dmimicro}
E_{\rm DM}=\sum_\alpha{\bf m}\cdot({\bf D}_\alpha\times\partial_\alpha{\bf m}),
\end{eqnarray}
where $\partial_\alpha=\partial/\partial \alpha$ is the spatial gradient along the direction ${\bf e}_\alpha$ and the DM vector ${\bf D}_\alpha$ fulfills Neumann's symmetry principle. As discussed in this work, one can show that ${\bf D}_\alpha$ possesses the same tensorial form as the current-driven damping-like torque tensor \cite{Freimuth2014}. From a theoretical standpoint, DMI is usually studied within either the atomistic or the micromagnetic limit. Whereas the atomistic form, Eq. \eqref{eq:dmiat}, is certainly more general, the micromagnetic form, Eq. \eqref{eq:dmimicro}, is often sufficient to describe the behavior of magnetic soft modes such as smooth domain walls and skyrmions. In contrast, the atomistic form is well adapted to study magnetic texture with strong, short-range canting like in weak ferromagnets and non-collinear antiferromagnets for instance.\par

The physical origin of this interaction at transition metal interfaces has been the object of numerous numerical investigations using density functional theory. The most straightforward approach consists in computing the energy of a spin cycloid or spiral in real space and determining the energy difference between states of opposite chirality. In density functional theory, such a spin spiral can be built by constraining the direction of the magnetic moments by applying a penalty energy on each of them \cite{Constrained}.Upon varying the length of the spin spiral (i.e., varying the size of the unit cell), the various DM vectors for nearest neighbors, next-nearest neighbors, etc. can be extracted using Eq. \eqref{eq:dmiat}. This approach has been used to compute the DM vector in ferroelectric magnets such as MgCr$_2$O$_4$\cite{Xiang2011} or Cu$_2$OSeO$_3$\cite{Yang2012b} and recently extended to transition metal interfaces\cite{Yang2015c}. The "constrained moment" method has the advantage of being applicable to materials with large spin-orbit coupling. However, it becomes computationally prohibitive in the long-wavelength limit (typically when the spin spiral wavelength exceeds 10 atomic sites) and is therefore more appropriate to compute the short-range DMI of insulating magnets than the long-range DMI of magnetic metals. \par

Alternatively, one can build spin spirals in the reciprocal space \cite{Kurz2004} employing the generalized Bloch theorem \cite{Herring1966,Sandratskii1991}. This approach, exact in the absence of spin-orbit coupling, permits the modeling of spin spirals of arbitrary wavelength. DMI is then computed to the {\em first order} in spin-orbit coupling \cite{Heide2008,Heide2009}. This method is limited to materials with weak enough spin-orbit coupling. DMI introduces an additional dispersion that is {\em odd} in the spin spiral momentum $q$ and the DM vector is usually evaluated taking the limit $q\rightarrow0$. This approach has been used to compute the DM vector in a wide range of transition metal interfaces\cite{Ferriani2008,Kashid2014,Zimmermann2014,Dupe2014,Schweflinghaus2016,Dupe2016,Belabbes2016,Belabbes2016b}. It is particularly well adapted to identify the emergence of chiral ground states, such as homochiral spin spirals \cite{Ferriani2008,Zimmermann2014}. \par

In the magnetic multilayers where N\'eel walls and room temperature skyrmions are observed, these chiral magnetic textures usually display smooth spatial gradient and long exchange length (typically 10 nm or more in perpendicularly magnetized materials). In this situation, the micromagnetic form, Eq. \eqref{eq:dmimicro}, seems more adapted to describe the onset of magnetic chirality. Within the micromagnetic limit, the DM vector can be computed by expanding the magnetic energy to the first order in magnetic gradient, an approach recently adopted by Freimuth et al. \cite{Freimuth2013b,Freimuth2014,Freimuth2017} and Kikuchi et al. \cite{Kikuchi2016}. Within linear response theory, it can be shown that the DM vector is related to the Berry curvature in the mixed spin-momentum space. In Kikuchi's theory, DMI is expressed as  $E_{\rm DM}=(\hbar/2)\int_\Omega {\bf m}\cdot[({\cal J}_s\cdot{\bm\nabla})\times{\bf m}]d^3{\bf r}$, where ${\cal J}_s$ is the equilibrium spin current that interacts with the magnetic texture. Mankovsky and Ebert\cite{Mankovsky2017} have recently computed DMI using Freimuth's theory implemented on fully relativistic Korringa-Kohn-Rostoker Green's function technique. \par

Irrespective of the method employed, the theoretical investigations of DMI at metallic interfaces have pointed out the importance of interfacial 3d-5d orbital hybridization \cite{Belabbes2016}. Since the magnetism is mostly localized on 3d orbitals whereas spin-orbit coupling is mostly carried by 5d orbitals, a proper balance between 3d and 5d orbitals is required to obtain large DMI, a trend confirmed experimentally \cite{Ma2018}. The role of orbital hybridization has also been indirectly probed through the dependence of DMI on the magnetization induced in the nonmagnetic metal \cite{Ryu2014,Rowan-Robinson2017}. While it is clear that DMI scales with 3d-5d hybridization \cite{Nembach2015}, the impact of inversion symmetry breaking on the magnitude of DMI has remained difficult to established experimentally. Recently, Kim et al. \cite{Kim2018a} demonstrated that DMI scales with the orbital asphericity arising from interfacial symmetry breaking, a feature confirmed by density functional theory.  This asphericity is associated with the equilibrium orbital magnetic moment, which was also suggested to play an important role in the onset of DMI \cite{Katsnelson2010}. We also recently proposed to tune DMI through interfacial oxidation \cite{Belabbes2016b}, an effect confirmed experimentally \cite{Chaves2019}.

A question that remains scarcely addressed is the localized or delocalized nature of DMI. For instance, considering magnetic transition metal chains deposited on top of nonmagnetic substrates, Kashid et al. \cite{Kashid2014} have pointed out that DMI extends far beyond the nearest neighbor interaction. Belabbes et al. \cite{Belabbes2016} showed that in W/Mn, DMI arises from the contribution of the first three W monolayers away from the interface. Experimentally, it is observed that DMI increases upon increasing the nonmagnetic metal thickness and saturates after a few nanometers \cite{Tacchi2017}, a scale that seems roughly comparable to the spin relaxation length. 

In the present work, we investigate the magnitude and symmetry of DMI in a nonmagnetic metal/ferromagnet heterostructure using a multi-orbital tight-binding model within the two-center Slater-Koster parameterization. We uncover the role of orbital mixing and show that DMI can extend over several monolayers away from the interface. Correspondingly, we examine the thickness dependence of DMI and find that it can be substantial. This Article is organized as follows: In Section \ref{s:keldysh}, we derive an expression for DMI to the first order in spatial gradient using Keldysh formalism. Then, Section \ref{s:model} presents the multi-orbital tight-binding model of the transition metal heterostructure. The results are discussed in Section \ref{s:results} and confronted to the oversimplified Rashba model. Finally, concluding remarks are given in Section \ref{s:conclusion}.

\section{Keldysh formalism for DMI\label{s:keldysh}}

As stated in the introduction, several methods have been proposed to compute DMI from first principles. To the best of our knowledge, the most popular approaches are the generalized Bloch theorem \cite{Heide2009} and the real-space spin spiral methods \cite{Yang2015}. In the present work, we aim to develop a Green's function formula that is suitable to our numerical platform. Such a Green's function formula has been derived by Freimuth et al. \cite{Freimuth2014} a few years ago by computing the energy of the system in the presence of a spin spiral and taking the long wave length limit. Here, we derive the DMI energy by computing the non-equilibrium response of the system in the presence of a gradient of magnetization within Keldysh formalism. As discussed below, in the limit of weak disorder and neglecting vertex corrections, our results boil down to the formula derived by Freimuth et al. \cite{Freimuth2014}.\par

Following Keldysh formalism \cite{Rammer1986,Onoda2008}, the lesser Green's function reads
\begin{eqnarray}
\hat{G}^<=(\hat{G}^R\otimes\hat{\Sigma}^<)\otimes\hat{G}^A,
\end{eqnarray}
where $\otimes\approx1+\frac{i\hbar}{2}\left(\overleftarrow{\partial}_{\bf p}\cdot\overrightarrow{\partial}_{\bf r}-\overleftarrow{\partial}_{\bf r}\cdot\overrightarrow{\partial}_{\bf p}\right)$ is the Moyal product expanded to the first order in spatial gradient. The retarded (advanced) Green's function fulfills Dyson's equation
\begin{eqnarray}\label{eq:dyson}
\left(\varepsilon-{\cal H}_0-\hat{\Sigma}^{R(A)}\right)\otimes\hat{G}^{R(A)}=\hat{1}.
\end{eqnarray} 
Here, ${\cal H}_0$ is the system's Hamiltonian in the absence of disorder, and the symbol $\overleftarrow{\partial}_i$ means that the derivative applies to the left, while $\overrightarrow{\partial}_i$ applies to the right. Let us now derive the lesser Green's function to the first order in spatial gradient. We obtain
\begin{eqnarray}\label{eq:gless}
\hat{G}^<&=&\hat{G}^R\hat{\Sigma}^<\hat{G}^A-\hbar{\rm Im}\left[\partial_{\bf p}\hat{G}^R\hat{\Sigma}^<\partial_{\bf r}\hat{G}^A\right]\\
&&-\hbar{\rm Im}\left[\partial_{\bf p}\hat{G}^R\partial_{\bf r}\hat{\Sigma}^<\hat{G}^A\right]+\hbar{\rm Im}\left[\partial_{\bf r}\hat{G}^R\partial_{\bf p}\hat{\Sigma}^<\hat{G}^A\right]\nonumber.
\end{eqnarray}
In the limit of short range impurities, the self-energies are local, i.e., $\hat{\Sigma}^{\alpha}=n_iV_0^2\int\frac{d^3{\bf k}}{(2\pi)^3}\hat{G}^\alpha$, with $\alpha=A,R,<$. Therefore, $\partial_{\bf p}\hat{\Sigma}^{\alpha}=0$ and the last term in Eq. \eqref{eq:gless} vanishes. Since the system is at equilibrium\cite{Rammer1986}, $\hat{\Sigma}^<=\left(\hat{\Sigma}^A-\hat{\Sigma}^R\right)f(\varepsilon)$, where $f(\varepsilon)$ is Fermi-Dirac distribution. Therefore, the first term in Eq. \eqref{eq:gless} reads
\begin{eqnarray}\label{eq:gless2}
\hat{G}^R\hat{\Sigma}^<\hat{G}^A=\hat{G}^R\left(\hat{\Sigma}^A-\hat{\Sigma}^R\right)\hat{G}^Af(\varepsilon).
\end{eqnarray}
This term must also be expanded to the first order in spatial gradient. To do so, one uses Dyson's equation for the retarded Green's function, Eq. \eqref{eq:dyson}, and expands the Moyal product. We obtain
\begin{eqnarray}\label{eq:dyson2}
\hat{G}^R&=&\hat{G}^R_0+\hat{G}^R_0\left(\hat{\Sigma}^R-\hat{\Sigma}_0^R\right)\hat{G}^R\hat{\Sigma}^<\hat{G}^A\\
&&-\frac{i\hbar}{2}\left(-\hat{G}^R_0\partial_{\bf p}{\cal H}_0\partial_{\bf r}\hat{G}^R+\hat{G}^R_0\partial_{\bf r}{\cal H}_0\partial_{\bf p}\hat{G}^R\right)\nonumber\\
&&-\frac{i\hbar}{2}\hat{G}^R_0\partial_{\bf r}\hat{\Sigma}^R\partial_{\bf p}\hat{G}^R.\nonumber
\end{eqnarray}
Here $\hat{\Sigma}_0^R$ is the self-energy at the zero-th order in spatial gradient, and we defined the unperturbed retarded Green's function $\hat{G}^R_0=\left(\varepsilon-{\cal H}_0-\hat{\Sigma}_0^R\right)^{-1}$. The first order perturbation of the retarded Green's function, $\hat{G}^R_{\bm\nabla}=\hat{G}^R-\hat{G}^R_0$, reads 
\begin{eqnarray}\label{eq:gR}
\hat{G}_{\bm\nabla}^R&=&\hat{G}^R_0\hat{\Sigma}_{\bm\nabla}^R\hat{G}^R_0\\
&&-\frac{i\hbar}{2}\left(\hat{G}^R_0\partial_{\bf r}{\cal H}_0\partial_{\bf p}\hat{G}^R_0-\hat{G}^R_0\partial_{\bf p}{\cal H}_0\partial_{\bf r}\hat{G}^R_0\right),\nonumber
\end{eqnarray}
where we defined $\hat{\Sigma}_{\bm\nabla}^R=\hat{\Sigma}^R-\hat{\Sigma}^R_0=n_iV_0^2\int\frac{d^3{\bf k}}{(2\pi)^3}\hat{G}^R_{\bm\nabla}$. After some algebra, and making use of 
\begin{eqnarray}
&&\hat{\Sigma}_0^A-\hat{\Sigma}_0^R=\left(\hat{G}_0^R\right)^{-1}-\left(\hat{G}_0^A\right)^{-1},\\
&&\partial_{\bf p}\hat{G}_0^R=\hat{G}_0^R\partial_{\bf p}{\cal H}_0\hat{G}_0^R,\\
&&\partial_{\bf r}\hat{G}_0^R=\hat{G}_0^R\partial_{\bf r}({\cal H}_0+\hat{\Sigma}_0^R)\hat{G}_0^R,
\end{eqnarray}
we obtain the final expression for the first order perturbation to the lesser Green's function, $\hat{G}_{\bm\nabla}^<={\rm Im}\left[\hat{G}_{\bm\nabla}^R\right]f(\varepsilon)$, where
\begin{eqnarray}\label{eq:ggrad}
\hat{G}_{\bm\nabla}^R&=&-\frac{i\hbar}{2}\hat{G}_0^R\left(\partial_{\bf r}({\cal H}_0+\hat{\Sigma}_0^R)\hat{G}_0^R\partial_{\bf p}{\cal H}_0\right.\\
&&\left.-\partial_{\bf p}{\cal H}_0\hat{G}_0^R({\cal H}_0+\hat{\Sigma}_0^R)+2i\hat{\Sigma}_{\bm\nabla}^R\right)\hat{G}_0^R.\nonumber
\end{eqnarray}
One notices that Eq. \eqref{eq:ggrad} involves self-consistent treatment of the disorder. In other words, $\hat{G}_{\bm\nabla}^R$ depends on $\hat{\Sigma}_{\bm\nabla}^R$, which shows that the above expression includes vertex corrections, in the same spirit as Ref. \onlinecite{Onoda2008}. Now, we can finally express the correction to the total energy
\begin{eqnarray}\label{eq:energy}
\langle{\cal H}_0-\mu\rangle=\hbar\int\frac{d\varepsilon}{2\pi i}{\rm Tr}\left[({\cal H}_0-\mu)\hat{G}_{\bm\nabla}^<\right].
\end{eqnarray}
By using the identity ${\cal H}_0-\mu=\varepsilon-\mu-\hat{\Sigma}_0^R-\left(\hat{G}^R_0\right)^{-1}$ and $-\left(\hat{G}_0^R\right)^2=\partial_\varepsilon\hat{G}_0^R$, we obtain the general expression for the DM energy
\begin{eqnarray}\label{eq:energy2}
&&\langle{\cal H}_0-\mu\rangle=A+B+C,
\end{eqnarray}
with
\begin{eqnarray}
A&=&-\hbar{\rm Re}\int\frac{d\varepsilon}{2\pi}(\varepsilon-\mu)f(\varepsilon){\rm Tr}\left[\left(\partial_{\bf r}({\cal H}_0+\hat{\Sigma}_0^R)\hat{G}_0^R\partial_{\bf p}{\cal H}_0\right.\right.\nonumber\\
&&\left.\left.-\partial_{\bf p}{\cal H}_0\hat{G}_0^R\partial_{\bf r}({\cal H}_0+\hat{\Sigma}_0^R)+2i\hat{\Sigma}_{\bm\nabla}^R\right)\partial_\varepsilon\hat{G}_0^R\right],\\
B&=&-\hbar{\rm Re}\int\frac{d\varepsilon}{2\pi}f(\varepsilon){\rm Tr}\left[\left(\partial_{\bf r}({\cal H}_0+\hat{\Sigma}_0^R)\hat{G}_0^R\partial_{\bf p}{\cal H}_0\right.\right.\nonumber\\
&&\left.\left.-\partial_{\bf p}{\cal H}_0\hat{G}_0^R\partial_{\bf r}({\cal H}_0+\hat{\Sigma}_0^R)\right)\hat{G}_0^R\hat{\Sigma}_0^R\hat{G}_0^R\right],\\
C&=&2\hbar{\rm Im}\int\frac{d\varepsilon}{2\pi}f(\varepsilon){\rm Tr}\left[\hat{\Sigma}_{\bm\nabla}^R\hat{G}_0^R\left(1+\hat{\Sigma}^R_0\hat{G}_0^R\right)\right].
\end{eqnarray}
Let us now simplify this formula. Neglecting the contribution of the self-energy, denoting $\hat{v}_j=\partial_{p_j}{\cal H}_0$ and recognizing that $\partial_{\bf r}{\cal H}_0=({\bf m}\times\partial_{\bf r}{\bf m})\cdot{\cal T}$, where ${\cal T}={\bf m}\times\partial_{\bf m}{\cal H}_0$ is the torque operator, we obtain
\begin{eqnarray}\label{eq:final}
\langle{\cal H}_0-\mu\rangle&=&\sum_{ij}D_{ij}{\bf e}_i\cdot({\bf m}\times\partial_j{\bf m})\\
D_{ij}&=&\hbar{\rm Re}\int\frac{d\varepsilon}{2\pi}(\varepsilon-\mu)f(\varepsilon)\times\nonumber\\
&&{\rm Tr}\left[{\cal T}_i\left(\partial_\varepsilon\hat{G}_0^R\hat{v}_j\hat{G}_0^R-\hat{G}_0^R\hat{v}_j\partial_\varepsilon\hat{G}_0^R\right)\right].\label{eq:dij}
\end{eqnarray}
This expression is exactly the one obtained in Ref. \onlinecite{Freimuth2014} (up to a "-" sign). This is the expression we will use in the next section to compute the DMI coefficient. 

\section{Tight-binding model\label{s:model}}

\subsection{Preliminaries}

\begin{figure}
\begin{center}
        \includegraphics[width=8cm]{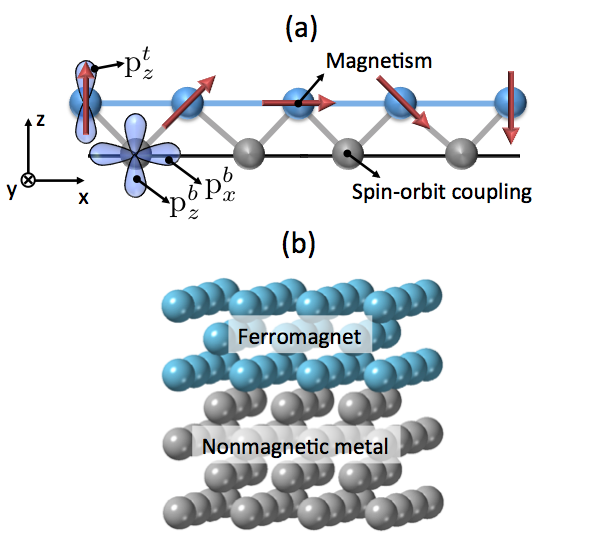}
      \caption{(Color online) Schematics of the two tight-binding models discussed in this work. (a) Two-orbital diatomic chain: The atoms of the bottom chain (gray) possess both p$_x$ and p$_z$ orbitals and spin-orbit coupling, while the atoms of the top chain (blue) has only p$_z$ orbitals and supports magnetism. (b) Multi-orbital bilayer heterostructure: The heterostructure is composed of two bcc monoatomic thin films whose elements possess all five d orbitals. The bottom layer (grey) is a nonmagnetic metal, whereas the top layer (blue) is magnetic. Both layers possess spin-orbit coupling.\label{fig:fig1}}
\end{center}
\end{figure}

Before entering into the details of the multi-orbital model proposed in this work, we introduce a simple minimal model for DMI, inspired from Ref. \onlinecite{Kashid2014}. The model is a diatomic chain along the $x$-direction, whose bottom non-magnetic atoms possess both p$_z$ and p$_x$ orbitals while the top magnetic atoms possess p$_z$ orbitals only. The bottom atoms possess spin-orbit coupling, while the top atoms carry magnetism. This toy model, depicted in Fig. \ref{fig:fig1}(a), represents an oversimplified nonmagnetic metal/ferromagnet heterostructure. In the $\{{\rm p}_z^t,{\rm p}_z^b,{\rm p}_x^b\}$ basis, where ${\rm p}_\nu^\eta$ is the $\nu$-th orbital of chain $\eta$, the Hamiltonian of the system reads

\begin{equation}\label{eq:ho}
{\cal H}_{\rm chain}=
\left(\begin{matrix}
\varepsilon_{k}^{t}& V_{zz} &V_{zx} \\
V_{zz}^* &\varepsilon_{k}^{z}&0\\
V_{zx}^*  &0&\varepsilon_{k}^{x}\\
\end{matrix}\right).
\end{equation}
Here p$_\nu^\eta$ refers to the $\nu$-th orbital of the top ($\eta=t$) or bottom chain ($\eta=b$), $V_{zz}=(V_\sigma+V_\pi)\cos k_xa/2$ and $V_{zx}=-i(V_\sigma-V_\pi)\sin k_xa/2$. $V_{\pi,\sigma}$ are the Slater-Koster hopping integrals\cite{Slater1954}. In addition, we turn on spin-orbit coupling ${\cal H}_{\rm so}$ on the bottom chain and magnetic exchange ${\cal H}_{\rm ex}$ on the top chain. Explicitly, 
\begin{equation}
{\cal H}_{\rm so}=\xi
\left(\begin{matrix}
0&0 &0 \\
0 &0&-i\hat\sigma_y\\
0 &i\hat\sigma_y&0\\
\end{matrix}\right),
\end{equation}
and 
\begin{equation}
{\cal H}_{\rm ex}=
\Delta\left(\begin{matrix}
\hat{\bm \sigma}\cdot{\bf m}&0 &0 \\
0 &0&0\\
0 &0&0\\
\end{matrix}\right).
\end{equation}
Let us now use Eq. \eqref{eq:final} to obtain an explicit expression of the DMI energy to the first order in exchange $\Delta$ and spin-orbit coupling $\xi$. By doing so, we intend to reveal the orbital mixing due to symmetry breaking that is at the origin of DMI. We first rewrite Eq. \eqref{eq:dij} as $D_{ij}=\hbar\int\frac{d\varepsilon}{2\pi}(\varepsilon-\mu)f(\varepsilon)g(\epsilon)$, with
\begin{eqnarray}\label{eq:ge}
g(\varepsilon)={\rm Re}{\rm Tr}\left[\hat{v}_j\hat{G}_0^R[\hat{G}_0^R,{\cal T}_i]\hat{G}_0^R\right].
\end{eqnarray}
The inner commutator can be extended to the first order in both spin-orbit coupling and exchange,
\begin{eqnarray}\label{eq:comm}
[\hat{G}_0^R,{\cal T}_i]\approx[\hat{G}_{00}^R{\cal H}_{\rm so}\hat{G}_{00}^R,{\cal T}_i],
\end{eqnarray}
where 
\begin{eqnarray}\label{eq:g00}
\hat{G}_{00}^R&=&(\varepsilon-{\cal H}_{0}+i0^+)^{-1},\\
&=&\sum_{n,s}\frac{|n\rangle\otimes|s\rangle\langle s|\otimes\langle n|}{\varepsilon-\varepsilon_n+i0^+},
\end{eqnarray}
and $|n\rangle\otimes|s\rangle$ is the eigenstate of ${\cal H}_{\rm chain}$, i.e., evaluated in the {\em absence} of spin-orbit coupling and exchange interaction. After some algebra, we obtain
\begin{eqnarray}
g(\varepsilon)=-\xi{\rm Re}\frac{{\rm Im}\left[\langle n|\hat{v}_j|m\rangle\langle m|{\cal T}_i|p\rangle\langle p|L_l|n\rangle\langle s|\sigma_k|s'\rangle\langle s'|\sigma_l|s\rangle\right]}{(\varepsilon-\varepsilon_m+i0^+)(\varepsilon-\varepsilon_p+i0^+)(\varepsilon-\varepsilon_n)^2}.\nonumber\\\label{eq:gen}
\end{eqnarray}
Summation over $n,m,p$ and $s,s'$ is assumed for short-handedness. The diagonalization of Hamiltonian \eqref{eq:ho} gives us three eigenstates. In order to make our result as simple as possible, we assume that $\varepsilon_{k}^{z}=\varepsilon_{k}^{x}$. Then, we end up with three bands with dispersion
\begin{eqnarray}
\varepsilon_{\bf k}^0&=&\varepsilon_{k}^{ z},\\
\varepsilon_{\bf k}^\pm&=&\frac{\varepsilon_{k}^{t}+\varepsilon_{k}^{z}}{2}\pm\frac{1}{2}\gamma_{\bf k},
\end{eqnarray}
with $\gamma_{\bf k}=\sqrt{(\varepsilon_{k}^{t}-\varepsilon_{k}^{z})^2+4(|V_{zz}|^2+|V_{zx}|^2)}$, corresponding to the eigenstates
\begin{eqnarray}\label{eq:wf}
|0\rangle&=&-\tilde{V}_{zx}|{\rm p}^b_z\rangle+\tilde{V}_{zz}|{\rm p}^b_x\rangle,\\
|+\rangle&=&\cos\chi|{\rm p}_z^t\rangle+\sin\chi\left(\tilde{V}_{zz}|{\rm p}_z^b\rangle+\tilde{V}_{zx}^*|{\rm p}_x^b\rangle\right),\\
|-\rangle&=&-\sin\chi|{\rm p}_z^t\rangle+\cos\chi\left(\tilde{V}_{zz}|{\rm p}_z^b\rangle+\tilde{V}_{zx}^*|{\rm p}_x^b\rangle\right),\nonumber\\
\end{eqnarray}
where $\cos2\chi=(\varepsilon_{\bf k}^t-\varepsilon_{\bf k}^z)/\gamma_{\bf k}$ and
\begin{eqnarray}
\tilde{V}_{zz}&=&\frac{V_{zz}}{\sqrt{|V_{zz}|^2+|V_{zx}|^2}},\;\tilde{V}_{zx}=\frac{V_{zz}}{\sqrt{|V_{zz}|^2+|V_{zx}|^2}}.\nonumber
\end{eqnarray}

\begin{figure}[h!]
        \includegraphics[width=4cm]{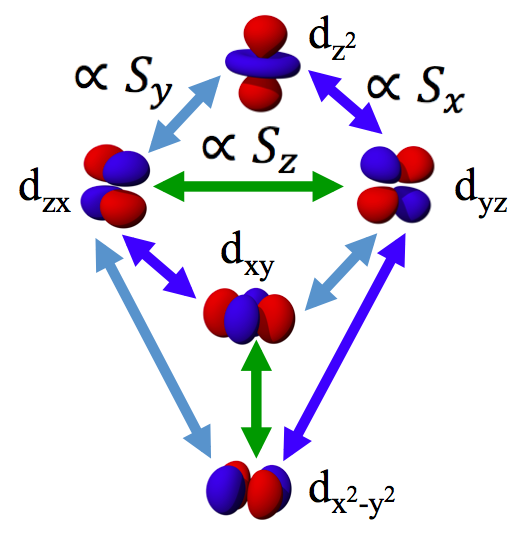}
      \caption{(Color online) Schematics of the spin momentum induced by mixing d orbitals in the presence of spin-orbit coupling.}
      \label{fig:fig2}
    \end{figure}

After some algebra, we obtain
\begin{eqnarray}\label{eq:gef}
D_{yx}&=&\hbar s\Delta\xi (V_\sigma^2-V_\pi^2)\int\frac{dk}{\Delta_k^5}\sin k_x a\times\\
&&\left[v_{x}^+f(\varepsilon_-)(\varepsilon_--\mu)-v_{x}^-f(\varepsilon_+)(\varepsilon_+-\mu)\right],\nonumber
\end{eqnarray}
and all the other matrix elements are zero. We retrieve in this simple expression all the key features of DMI at interfaces. It is, to the lowest order, linear in both spin-orbit coupling and magnetic exchange and proportional to the inversion symmetry breaking through $V_{zx}$. This potential characterizes the admixture between p$_z^b$ and p$_x^b$ orbitals, mediated by p$_z^t$ orbitals. This admixture enables the onset of an orbital momentum along ${\bf y}$, which results in the emergence $D_{yx}$. One can extend this scenario to d orbitals: admixture between two orthogonal orbitals, mediated by a symmetry breaking coupling term, can result in a non-vanishing orbital momentum, as illustrated in Fig. \ref{fig:fig2}. The multi-orbital tight-binding model presented below intends to encompass such admixtures at interfaces. \par

\subsection{Transition metal heterostructure}

We now move on to the description of the tight-binding model of our transition metal heterostructure. Since this method has been described in Ref. \onlinecite{Manchon2019c}, we summarize its main features below and refer the reader to Ref. \onlinecite{Manchon2019c} for more details. The structure is depicted on Fig. \ref{fig:fig1}(b) and consists of two adjacent transition metal layers with bcc crystal structure and equal lattice parameter. The model is constituted of monolayers stacked on top of each other along the (001) direction. The individual Hamiltonian of a monolayer reads
\begin{equation}
{\cal H}_{0}={\cal H}_{\rm mono}\otimes\hat{\sigma}_0+{\cal H}_{\rm ex}+{\cal H}_{\rm soc}.
\end{equation}
The first term is the 10$\times$10 Hamiltonian of the monolayer without magnetic exchange. ${\cal H}_{\rm mono}$ is written in the basis $\{{\rm d}_{xy},{\rm d}_{yz},{\rm d}_{zx},{\rm d}_{x^2-y^2},{\rm d}_{z^2}\}$ and its matrix elements are written assuming two-center Slater-Koster parameterization \cite{Slater1954}. The second term is the exchange interaction between the itinerant spins and the magnetic order, and the third term is the spin-orbit coupling Hamiltonian written in Russel-Saunders scheme, 
\begin{equation}
{\cal H}_{\rm soc}=\xi_{\rm so}
\left(\begin{matrix}
0&i\hat{\sigma}_y &-i\hat{\sigma}_x  &0&2i\hat{\sigma}_z\\
-i\hat{\sigma}_y&0&i\hat{\sigma}_z &-i\sqrt{3}\hat{\sigma}_x &-i\hat{\sigma}_x\\
i\hat{\sigma}_x  &-i\hat{\sigma}_z&0&i\sqrt{3}\hat{\sigma}_y &-i\hat{\sigma}_y\\
0&i\sqrt{3}\hat{\sigma}_x&-i\sqrt{3}\hat{\sigma}_y &0&0\\
-2i\hat{\sigma}_z&i\hat{\sigma}_x&i\hat{\sigma}_y&0&0\\
\end{matrix}\right).\label{eq:soc}
\end{equation}

Each monolayer is connected to its top first and second-nearest neighbor through off-diagonal matrices, ${\cal T}_1$ and ${\cal T}_2$, respectively. The Hamiltonian of one bcc layer is then 
\begin{equation}
\cal{H}_{\rm layer}=
\left(\begin{matrix}
{\cal H}_0 & {\cal T}_1 & {\cal T}_2 & 0&\\
{\cal T}^\dagger_1 & {\cal H}_0& {\cal T}_1 & {\cal T}_2  &\ddots\\
{\cal T}^\dagger_2 &{\cal T}^\dagger_1& {\cal H}_0&{\cal T}_1&\ddots \\
0&{\cal T}^\dagger_2 &{\cal T}^\dagger_1&{\cal H}_0&\ddots\\
& \ddots&\ddots& \ddots&\ddots\\
\end{matrix}\right)
\end{equation}

%\begin{table}
%\begin{tabular}{c|cccccc}
%& $V_\sigma^1$ & $V_\pi^1$ & $V_\delta^1$& $V_\sigma^{2}$ & $V_\pi^2$ & $V_\delta^2$\\\hline
%FM & -0.618 & 0.37 &-0.035&-0.37&0.08 & 0.01\\
%NM & -1.61 & 0.71 &0.034&-0.99 & -0.17 & 0.12\\
%FM/NM & -1.11 & 0.54 &-0.001 & -0.68 & -0.046 &0.066\\
%\end{tabular}
%\begin{tabular}{c|ccccc}
%&$\varepsilon_{t_{2g}}$&$\varepsilon_{e_g}$&$\Delta_{t_{2g}}$&$\Delta_{e_{g}}$&$\xi_{\rm soc}$\\\hline
%FM & 12.8 & 12.5 &1.85&1.73&0.065 \\
%NM & 13.16 & 11.66 & 0 &0 & 0.367  \\
%\end{tabular}
%\caption{Slater-Koster parameters used to model the FM/NM heterostructure. The parameters are given in eV.\label{table1}}
%\end{table} 

%The Slater-Koster parameters of each layers are given in Table \ref{table1}. 
We adopt the parameters computed by Papaconstantopoulos \cite{Papaconstantopoulos2015} for {\em bulk} bcc Fe and bcc W (see Ref. \cite{Manchon2019c} for details). With these parameters, we determine the Hamiltonian for the nonmagnetic and ferromagnetic layers, $\cal{H}_{\rm NM}$ and $\cal{H}_{\rm F}$. Finally, the heterostructure is obtained by stitching two individual slabs together, yielding the total Hamiltonian
\begin{equation}
\cal{H}=
\left(\begin{matrix}
{\cal H}_{\rm F} & {\cal T}^{\rm FN} \\
{\cal T}^{\rm FN,\dagger} & {\cal H}_{\rm NM}\\
\end{matrix}\right).
\end{equation}

\begin{figure}[h!]
 \centering
        \includegraphics[width=7cm]{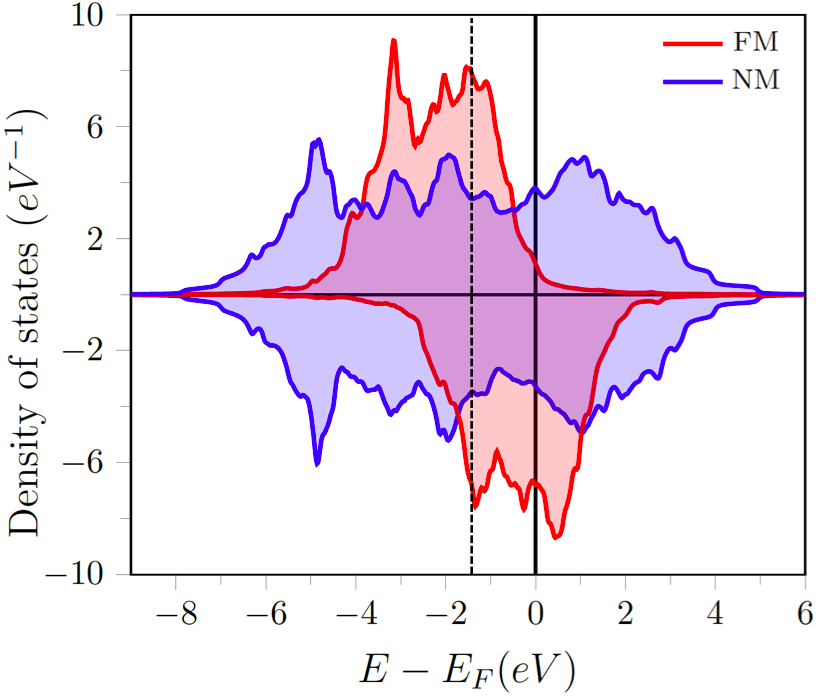}
        \caption{(Color online) The spin-resolved density of states of FM(5)/NM(7) bilayer with $E_F=14$ eV. The blue shaded area corresponds to the contribution of the nonmagnetic metal, while the red shaded area corresponds to the ferromagnetic metal contribution. The vertical dotted line indicates $E_F=12.6$ eV.}
        \label{fig:fig3}
    \end{figure}

The hopping matrix ${\cal T}^{\rm FN}$ is simply given by ${\cal T}_1$ and ${\cal T}_2$ adopting the parameters of Table I in Ref. \cite{Manchon2020}. At zero temperature, the chemical potential equals the Fermi energy, $\mu=E_{\rm F}=14$ eV. The density of state of the structure can be obtained by computing $-\frac{1}{\pi}{\rm Im}[\hat{G}^R]$, where $\hat{G}^R=(\varepsilon-{\cal H}+i\Gamma)$ is the retarded Green's function and $\Gamma$ is the homogeneous broadening, as shown in Fig. \ref{fig:fig3}. Our minimal multi-orbital model serves as a platform to our investigation on DMI.\par

We would like to emphasize that because magnetism arises from both spin and orbital moments, DMI also possesses both orbital and spin contributions, as discussed in the case of La$_2$CuO$_4$ by Ref. \onlinecite{Katsnelson2010}. In certain systems, such as correlated oxides, the orbital contribution to the overall magnetism is important and therefore can substantially contribute to DMI\cite{Katsnelson2010}. The theory presented in section \ref{s:keldysh} does in principle account for the orbital contribution. Nonetheless, in transition metal multilayers the orbital moment is usually quenched due to the high symmetry of the bulk metal and slightly increases close to the interface due to symmetry lowering \cite{Grytsyuk2016}. Although we acknowledge that the influence of this orbital moment deserves further study, we neglect this orbital contribution in the present work.

\begin{widetext}
    \begin{figure*}[t]
\centering
        \includegraphics[width=18cm]{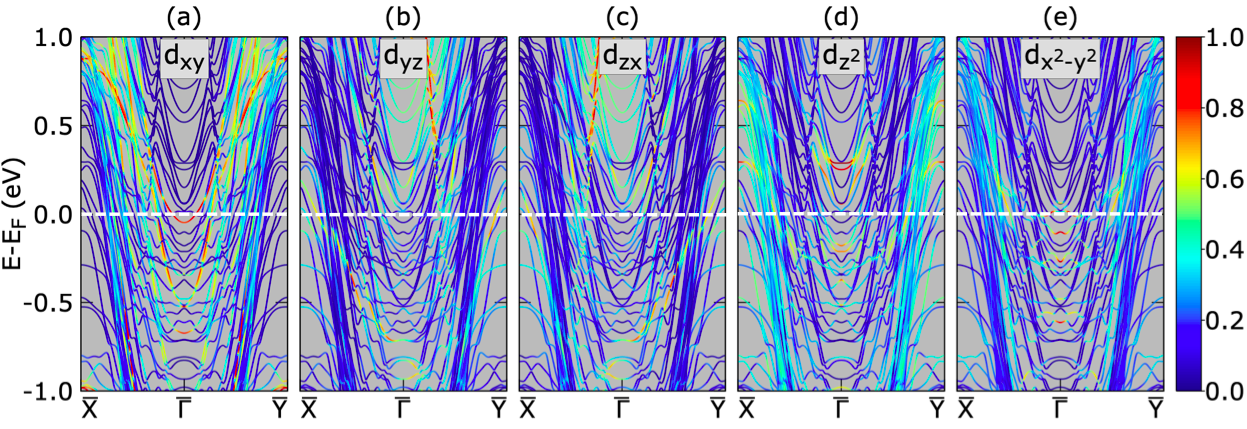}
      \caption{(Color online) Orbital-resolved band structure of FM(5)/NM(7) bilayer. The band structure is projected on (a) d$_{xy}$, (b) d$_{yz}$, (c) d$_{zx}$, (d) d$_{z^2}$ and (e) d$_{x^2-y^2}$. The scale spans from 0 (dark blue) to 1 (red). Close to Fermi level (white dashed line), one can see that d$_{xy}$, d$_{z^2}$ and d$_{x^2-y^2}$ contributions are isotropic in momentum, while d$_{yz}$ and d$_{zx}$ contributions are anisotropic.}
      \label{fig:fig4}
    \end{figure*}
    \end{widetext}

Before closing this brief presentation, let us inspect the band structure of the heterostructure along the $\bar{\rm X}-\bar{\Gamma}-\bar{\rm Y}$ path, projected on the various d orbitals, as displayed in Fig. \ref{fig:fig4}. As mentioned in the previous section, the admixture of two orthogonal such orbitals favors the onset of DMI (see Fig. \ref{fig:fig2}), and it is therefore instructive to identify the momentum-dependent orbital texture close to Fermi level. From Fig. \ref{fig:fig4}, we see that d$_{xy}$, d$_{x^2-y^2}$ and d$_{z^2}$ are isotropic in momentum [Figs. \ref{fig:fig4}(a), (d), and (e)], d$_{z^2}$ being dominant [light blue in Fig. \ref{fig:fig4}(d)] over d$_{xy}$ and d$_{x^2-y^2}$  [blue in Figs. \ref{fig:fig4}(a,e)] at Fermi level. In contrast, d$_{yz}$ and d$_{zx}$ are weaker [dark blue in Figs. \ref{fig:fig4}(b,c)] and display an anisotropic texture: their magnitude along along the $\bar{\Gamma}-\bar{\rm Y}$ path is different from their magnitude along $\bar{\Gamma}-\bar{\rm X}$ path.

\begin{figure*}[t]
\centering
        \includegraphics[width=12cm]{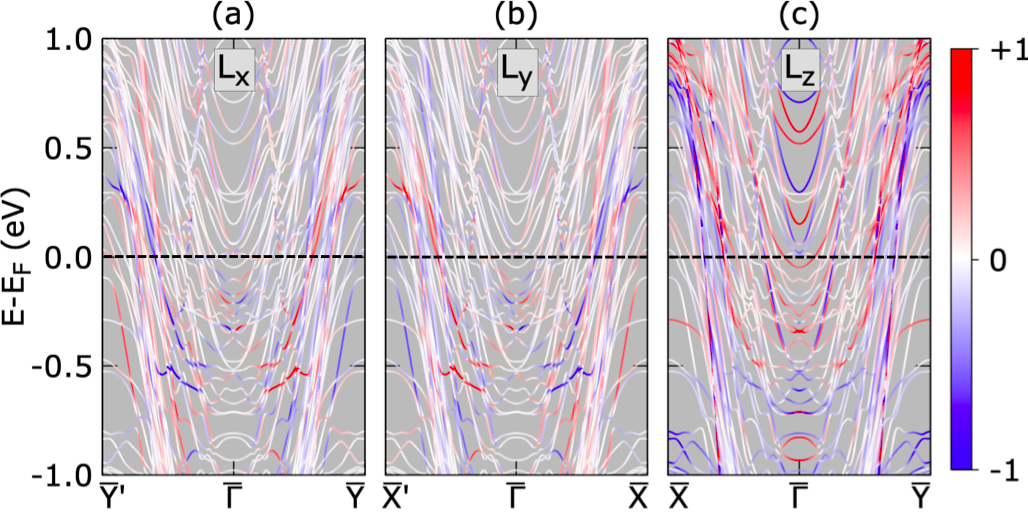}
      \caption{(Color online) Band structure of FM(5)/NM(7) bilayer projected on the three components of the orbital momentum. $L_x$ and $L_y$ are antisymmetric along $\bar{\rm Y}-\bar{\Gamma}-\bar{\rm Y}$ and $\bar{\rm X}-\bar{\Gamma}-\bar{\rm X}$, respectively, whereas $L_z$ remains even in momentum.\label{fig:fig4bis}}
    \end{figure*}

This feature promotes the inverse orbital galvanic effect, i.e., the generation of non-equilibrium orbital momentum \cite{Yoda2018}, illustrated on Fig. \ref{fig:fig4bis}. From Fig. \ref{fig:fig4bis}(a,b), we see that the $L_x$ and $L_y$ components are antisymmetric in momentum $k$ along $\bar{\Gamma}-\bar{\rm Y}$ and $\bar{\Gamma}-\bar{\rm X}$ paths, respectively. In contrast, the $L_z$ component is isotropic and {\em even} in momentum. In other words, ${\bf L}\propto {\bf z}\times{\bf k}$. As a consequence, based on Fig. \ref{fig:fig2} and Fig. \ref{fig:fig4}, we can propose the following scenario: around Fermi level, the admixture d$_{xy}$-d$_{zx}$ (d$_{xy}$-d$_{yz}$) produces a non-equilibrium orbital momentum $L_y$ ($L_x$) along the $\bar{\Gamma}-\bar{\rm X}$ ($\bar{\Gamma}-\bar{\rm Y}$) path. In the presence of magnetization gradient along ${\bf x}$, this orbital momentum promotes the onset of a DM vector along ${\bf y}$, i.e., $D_{yx}$. Symmetrically, the admixture d$_{xy}$-d$_{yz}$ promotes the onset of a DM vector along ${\bf x}$ for a magnetization gradient along ${\bf y}$, i.e., $D_{xy}$.

\section{Results\label{s:results}}

\begin{figure}[h!]
        \includegraphics[width=8cm]{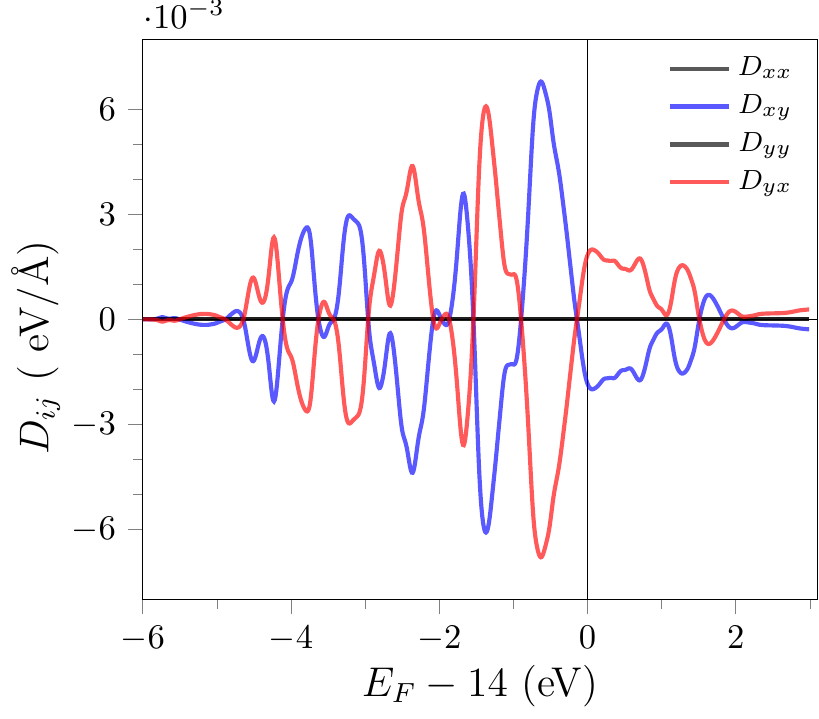}
      \caption{(Color online) The different coefficients of the DMI tensor as a function of the Fermi energy, computed in FM(5)/NM(7) bilayer using our multi-orbital tight-binding approach. In this calculation, the broadening is set to $\Gamma=50$ meV and $E_{\rm F}=14$ eV.}
      \label{fig:fig5}
    \end{figure}
\subsection{Symmetry analysis}

As explicitly demonstrated in Ref. \onlinecite{Freimuth2014}, the DMI coefficient $D_{ij}$ possesses the same symmetries as the damping-like torque coefficient, $t_{ij}$, defined as $T_{\rm DL}^i=t_{ij}E_j$, $E_{j}$ being the $j$-th component of the electric field. At an interface with the highest symmetry ${\cal C}_\infty$, the damping-like torque reads\cite{Manchon2019}
\begin{eqnarray}\label{eq:dampingtorque}
{\bf T}_{\rm DL}\propto {\bf m}\times[{\bf m}\times({\bf z}\times{\bf E})],
\end{eqnarray}
or, equivalently,
\begin{equation}
\hat{t}_{{\cal C}_\infty}\propto\left(\begin{matrix}
m_xm_y&m_z^2+m_y^2 &0 \\
-m_z^2-m_y^2&-m_xm_y&0\\
m_ym_z &-m_xm_z&0\\
\end{matrix}\right)\equiv\hat{D}_{{\cal C}_\infty}.
\end{equation}
By identifying the matrix elements of $\hat{D}_{{\cal C}_\infty}$ to that of $\hat{t}_{{\cal C}_\infty}$, we obtain the DMI energy
\begin{equation}
E_{\rm DM}=D{\bf m}\cdot[({\bf z}\times{\bm\nabla})\times{\bf m}],
\end{equation}
as expected at such interfaces. The coefficient $D$ can be obtained by solving Eq. \eqref{eq:final} for the magnetic Rashba Hamiltonian,
\begin{equation}\label{eq:rashba}
{\cal H}=\frac{\hat{\bf{p}}^{2}}{2 m}+\alpha_{\rm R}(\mathbf{z} \times \hat{\boldsymbol{\sigma}}) \cdot {\bf k}+\Delta \hat{\boldsymbol{\sigma}} \cdot {\bf m}.
\end{equation}
In the constant relaxation time approximation, the retarded Green's function reads $\hat{G}^R=\frac{1}{2}\sum_s\frac{1+s\hat{\bm \sigma}\cdot{\bf n}}{\varepsilon-\varepsilon_{k,s}+i\Gamma}$, where
\begin{eqnarray}
\varepsilon_{k,s}&=&\frac{\hbar^2k^2}{2m}+s\lambda_k,\;{\bf n}=-(\alpha_{\rm R}/\lambda_k)\mathbf{z} \times {\bf k},\\
\lambda_k&=&\sqrt{\Delta^2+\alpha_{\rm R}^2-2\Delta\alpha_{\rm R}k\sin\theta\sin(\varphi-\varphi_m)}.
\end{eqnarray}
Using Eq. \eqref{eq:final}, we obtain
\begin{eqnarray}
E_{\rm DM}&=&-\hbar \Delta{\rm Re}\int\frac{d\varepsilon}{2\pi}(\varepsilon-\varepsilon_F)f(\varepsilon)\int\frac{d^2{\bf k}}{4\pi^2}\times\\
&&{\rm Tr}\left[\hat{v}_j\hat{G}^R(\hat{\bm\sigma}\cdot\partial_j{\bf m})\partial_\varepsilon \hat{G}^R-\hat{v}_j\partial_\varepsilon \hat{G}^R(\hat{\bm\sigma}\cdot\partial_j{\bf m}) \hat{G}^R\right].\nonumber\\
&=&\alpha_{\rm R}\Delta\sum_{s}\int\frac{d^2{\bf k}}{2\pi^2}\frac{s(\varepsilon_{k,s}-E_F)f(\varepsilon_{k,s})}{(\varepsilon_{k,s}-\varepsilon_{k,-s})^2}[{\bf n}\times({\bf z}\times{\bm\nabla})]\cdot{\bf m}\nonumber\\
\end{eqnarray}
After solving the integral, we get the final expression
\begin{equation}\label{dmirashba}
E_{\rm DM}=\frac{\alpha_{\rm R} E_F}{2 \pi}\frac{m}{ \hbar^{2}} {\bf m} \cdot\left[\left({\bf z} \times {\bm \nabla}\right) \times {\bf m}\right],
\end{equation}
from which we can see that
\begin{eqnarray}
&&D_{x y}=-D_{y x}=\frac{\alpha_{\rm R} E_F}{2 \pi}\frac{m}{ \hbar^{2}},\\
&&D_{x x}=D_{y y}=0
\end{eqnarray}
In this expression, the magnetic exchange $\Delta$ does not appear explicitly due to an accidental cancellation with the denominator $\propto \varepsilon_{k,s}-\varepsilon_{k,-s}$, a feature specific to the ideal case of the free electron Rashba gas. 

In systems that deviate from the ideal Rashba case, such as transition metal multilayers, the damping-like torque display higher order behavior that feature torque components of unusual symmetry \cite{Belashchenko2019,Belashchenko2020}, beyond that of Eq. \eqref{eq:dampingtorque}. Since we did observe these additional features in the tight-binding model presented here\cite{Manchon2020}, one could reasonably expect that the associated DMI might also display higher order contributions. The DMI coefficients for the transition metal heterostructure are reported on Fig. \ref{fig:fig5} as a function of the Fermi energy. The multi-orbital tight-binding model, in spite of its much higher complexity than the Rashba Hamiltonian, Eq. \eqref{eq:rashba}, also displays $D_{xx}=D_{yy}=0$ and $D_{xy}=-D_{yx}\neq0$, in contrast with the damping-like torque discussed above.\par

\begin{figure}[h!]
        \includegraphics[width=8cm]{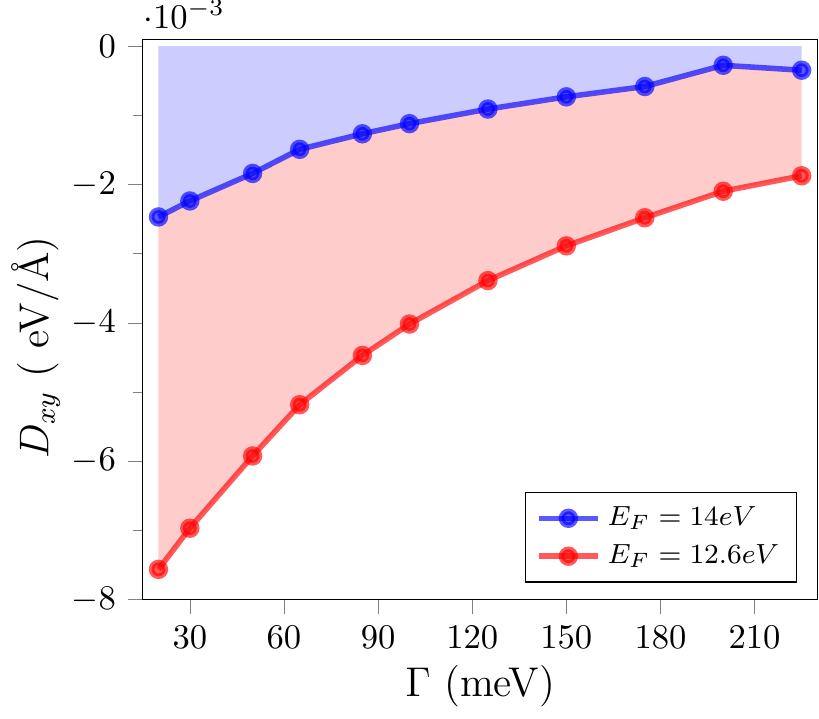}
      \caption{(Color online) The DMI coefficient $D_{xy}$ as a function of the disorder broadening in FM(5)/NM(7) bilayer, for two values of the Fermi energy.}
      \label{fig:fig6}
    \end{figure}

We conclude this discussion by computing the $D_{xy}$ coefficient as a function of the disorder strength $\Gamma$. As exposed in Eqs. \eqref{eq:gef} and \eqref{dmirashba}, DMI is an intrinsic mechanism in the sense that when disorder vanishes, it converges to a finite value \cite{Freimuth2014}, similarly to the damping-like torque in this respect \cite{Freimuth2014a}.  However, as discussed in Section \ref{s:keldysh}, in the presence of short-range disorder the self-energy is non-vanishing and reads $\hat{\Sigma}^{R(A)}=n_iV_0^2\int\frac{d^3{\bf k}}{(2\pi)^3}\hat{G}^{R(A)}$. This self-energy should be computed self-consistently in order to account for all scattering orders. Since this procedure is highly computationally demanding, it is conventional to reduce the self-energy to a constant broadening $\hat{\Sigma}^{R(A)}=\mp i\Gamma$, equivalent to the constant relaxation time approximation in Boltzmann transport equation. We report the disorder dependence of the $D_{xy}$ coefficient for two different Fermi energies in Fig. \ref{fig:fig6}. Due to numerical limitations, we could not test the limit of vanishing disorder. Nonetheless, the DMI coefficient displays a smooth decay as a function of disorder [$\propto 1/(\eta^2+\Gamma^2)$], smaller than the one expected for an extrinsic effect ($\propto 1/\Gamma$). This observation is important because it emphasizes the major impact of disorder on the DMI coefficient, despite its "intrinsic" origin. In the ideally clean limit, the intrinsic origin of the DMI reveals itself through the importance of the "band anticrossing", resulting in sharp peaks when spanning across the band structure \cite{Koretsune2015,Sandratskii2017}. In real materials though, thermally activated phonons and defects induce a finite broadening $\Gamma$, which washes out these singularities. From the transport calculations performed in Ref. \onlinecite{Manchon2019c}, we estimate that this broadening is about 20 meV, corresponding to a conductivity of $\sim$10$^7$ $\Omega^{-1}\cdot$m$^{-1}$. In other words, in realistic systems the value of DMI is unlikely to be equal to the one obtained in the clean limit and should be substantially smaller. An estimation solely based on the clean limit systematically overestimates DMI.

\subsection{Orbital decomposition of DMI}

As we have seen in Section \ref{s:model}, DMI arises from the orbital momentum stemming from the admixture of the atomic orbitals induced by symmetry breaking. Whereas the toy model of Section \ref{s:model} was based solely on p$_x$ and p$_z$ orbitals, giving rise to $L_y$ orbital momentum, our multi-orbital model for the transition metal heterostructure involves all the 10 d-orbitals. To understand which orbitals are involved in the emergence of interfacial DMI, one can contemplate the chart provided in Fig. \ref{fig:fig2}. This figure schematically represents the spin momentum direction induced by the atomic spin-orbit coupling upon the mixing of two d atomic orbitals. \par

\begin{figure}[h!]
        \includegraphics[width=8cm]{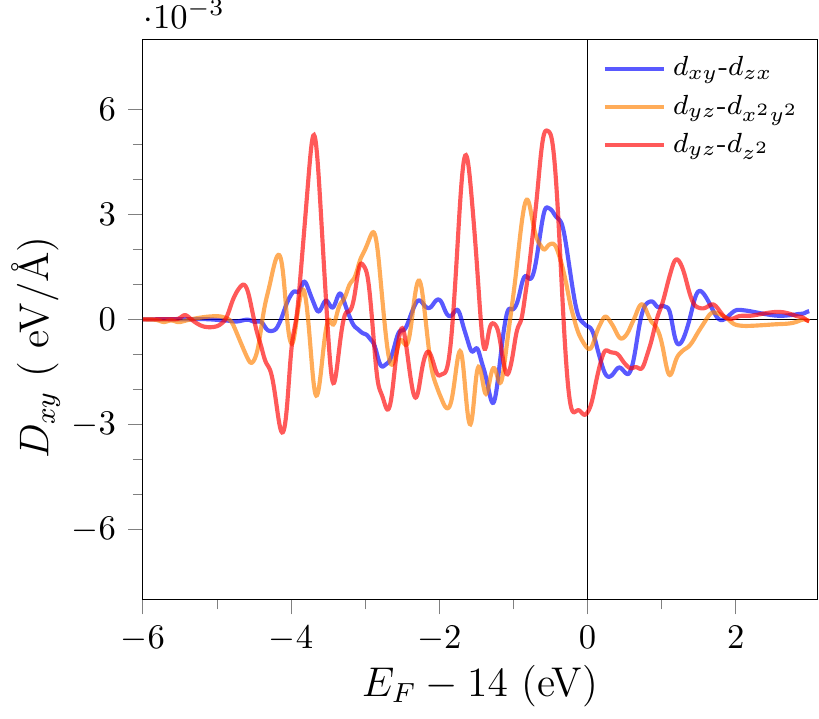}
      \caption{\label{fig:fig7} (Color online) $D_{xy}$ as a function of the Fermi energy when turning on only specific coefficients of the spin-orbit coupling matrix, in FM(5)/NM(7) bilayer. Here the spin-orbit coupling of the ferromagnetic layer is turned off. The broadening is set to $\Gamma=50$ meV and $E_{\rm F}=14$ eV.}
    \end{figure}

In order to stabilize a perpendicular N\'eel spin spiral propagating along ${\bf x}$, the orbital momentum must be aligned along ${\bf y}$, which can be obtained by the following admixture: d$_{zx}$-d$_{z^2}$, d$_{xy}$-d$_{yz}$ and d$_{zx}$-d$_{x^2-y^2}$. Similarly, in order to induce a perpendicular N\'eel spin spiral along ${\bf y}$, the orbital momentum must be aligned along ${\bf x}$, which can be obtained by mixing: d$_{yz}$-d$_{z^2}$, d$_{zx}$-d$_{xy}$ and d$_{yz}$-d$_{x^2-y^2}$. In Fig. \ref{fig:fig7}, the DMI coefficient $D_{xy}$ is calculated by only turning on the spin-orbit coupling coefficient that mixes two specific orbitals. In this figure, the spin-orbit coupling of the ferromagnetic layer is set to zero, for simplicity. We see that the dominant contributions to DMI come from d$_{yz}$-d$_{z^2}$ (red), d$_{zx}$-d$_{xy}$ (cyan) and d$_{yz}$-d$_{x^2-y^2}$ (orange), all orbital combinations giving an orbital momentum along $L_x$. We also notice the reduced contribution from d$_{xy}$-d$_{x^2-y^2}$ (blue), which produces an orbital momentum along $L_z$. This orbital-resolved diagram demonstrates that the scenario discussed in Section \ref{s:model} remains mostly valid in our multi-orbital system. Notice that the specific orbital contributions are strongly energy dependent, which reflects the fact that the electronic band structure displays strong orbital texture [Fig. \ref{fig:fig4}]. Finally, we emphasize that performing the same analysis on the DMI coefficient $D_{yx}$ gives orbitals combinations that yield an orbital momentum $L_y$.

\begin{figure}[h!]
        \includegraphics[width=8cm]{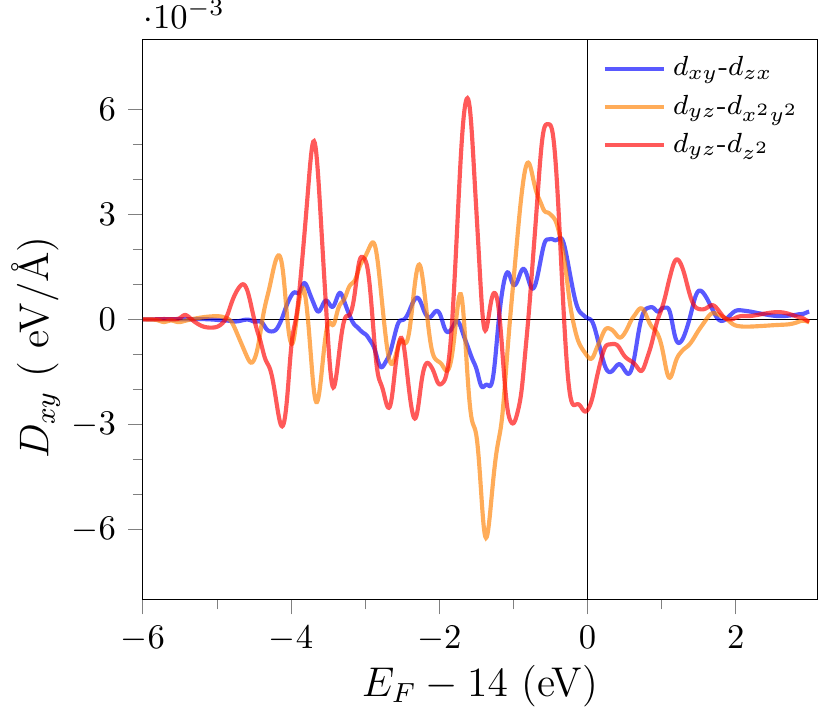}
      \caption{\label{fig:fig8} (Color online) $D_{xy}$ as a function of the Fermi energy when turning on only specific coefficients of the spin-orbit coupling matrix, in FM(5)/NM(7) bilayer. The spin-orbit coupling of both the ferromagnetic and nonmagnetic layers is turned on. The broadening is set to $\Gamma=50$ meV and $E_{\rm F}=14$ eV.}
    \end{figure}

Figure \ref{fig:fig8} displays the same orbital-resolved DMI when spin-orbit coupling is present in both metals. Whereas the DMI orbital decomposition remains mostly unaffected for low ($E_F<12$ eV) and high energies ($E_F>15$ eV), we notice that the contribution from d$_{xy}$-d$_{x^2-y^2}$ (blue) increases substantially, reflecting the important role of interfacial orbital mixing between the magnetic and nonmagnetic orbitals.

\subsection{Thickness dependence}

An important question that remained to be addressed is whether DMI is localized at the interface or whether it extends away from it. As mentioned in the introduction, it has been experimentally observed that in CoFeB/Pt heterostructure DMI increases upon increasing the nonmagnetic metal thickness and saturates after a few nanometers \cite{Tacchi2017}, on a scale that seems roughly comparable to the spin relaxation length. From the theoretical viewpoint, Yang et al. \cite{Yang2015c} computed DMI in Co/Pt(111) and found that is it dominated by the uppermost Pt layer, while Belabbes et al. \cite{Belabbes2016} computed DMI in Mn/W(001) and found that the first three W layers contribute to the total DMI. Although these two calculations are performed using different methods (real-space spin spiral versus momentum-space spin spiral), they indicate that different materials may display quite different behaviors. \par

\begin{figure}[h!]
        \includegraphics[width=8cm]{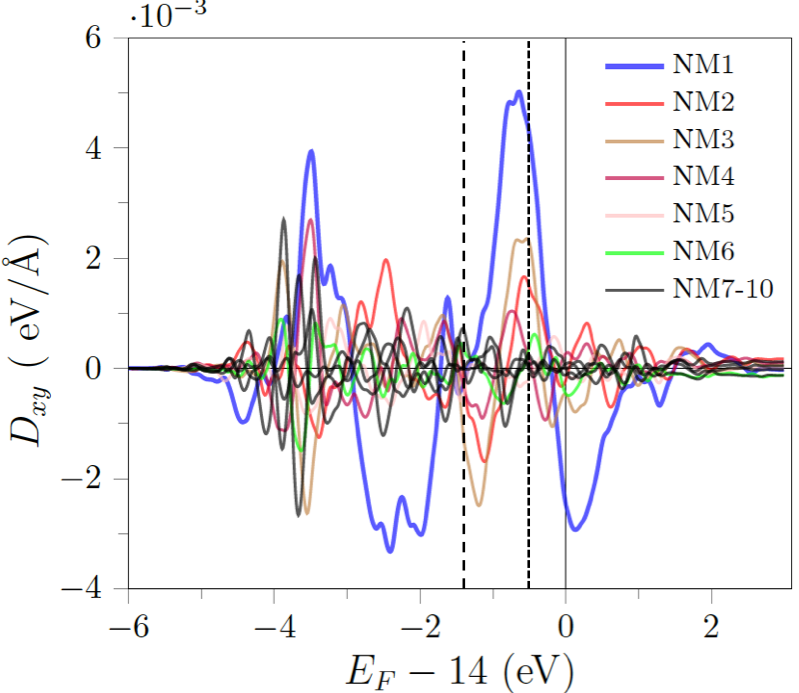}
      \caption{(Color online) $D_{xy}$ as a function of the Fermi energy when turning on the spin-orbit coupling only on specific layers. The vertical dashed and dotted lines indicate $E_F=12.6$ eV and $E_F=13.5$ eV, respectively.}
      \label{fig:fig9}
    \end{figure}

In Fig. \ref{fig:fig9}, we report the energy-dependent DMI coefficient when turning on the spin-orbit coupling parameter of a given monolayer away from the interface in FM(3)/NM(10)  while turning off the spin-orbit coupling of the other layers. This procedure is only valid in the limit of weak spin-orbit coupling, but does provide a qualitative picture of the delocalized nature of DMI as long as the overall band structure remains weakly modified by the spin-orbit coupling of individual layers. Figure \ref{fig:fig9} shows that whereas DMI is often dominated by the uppermost nonmagnetic metal monolayer (thick blue line), the contribution of the sub-monolayers is very sensitive to the energy (thin colored lines). At Fermi energy, DMI is entirely dominated by the uppermost nonmagnetic metal layer. However, around 13.5 eV contributions from the second and third monolayers become significant (vertical dotted line in Fig. \ref{fig:fig9}), indicating that the Bloch states participating to DMI have a delocalized character. At 12.6 eV, only the second and third layers contribute whereas the first layer close to the interface does not (vertical dashed line in Fig. \ref{fig:fig9}). This complex behavior reflects again the high sensitivity of the orbital composition of the band structure as a function of the energy. It also indicates that the nature of DMI, localized close to the interface or delocalized away from it, is material sensitive. This suggests that such a feature could be tuned by doping the nonmagnetic metal and modifying the Fermi level.\par

    %12.6eV which is  the state that optimizes the system to yield the Maximum DMI by having enough down states of Fe and also W states acting as the channel of mixing between the up and down states.%

\begin{figure}[h!]
        \includegraphics[width=8cm]{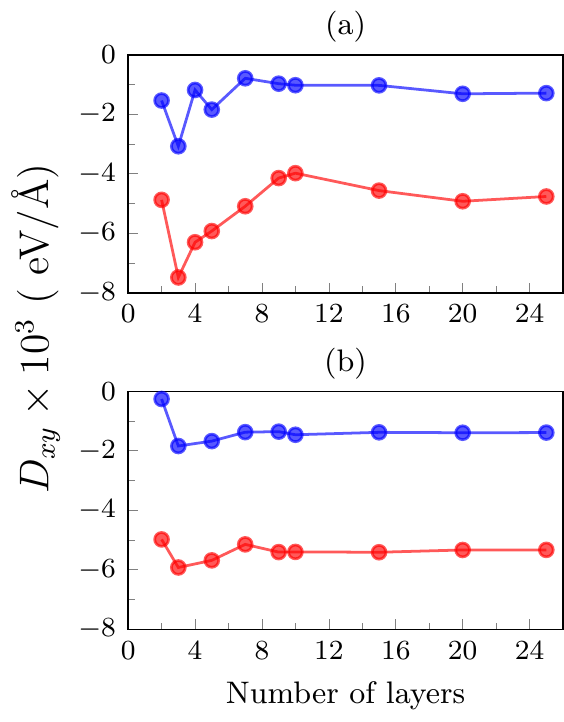}
      \caption{(Color online) Thickness dependence of $D_{xy}$ when (a) varying the nonmagnetic thickness and setting the ferromagnetic layer to 3 monolayers, and (b) varying the ferromagnetic thickness and setting the nonmagnetic layer to 5 monolayers. The calculations have been performed for $E_F=14$ eV (blue) and 12.6 eV (red).}
      \label{fig:fig10}
    \end{figure}

To conclude this study, let us now turn our attention towards the thickness dependence of DMI. Upon varying the thickness of the nonmagnetic metal, DMI shows a large modulation as reported in Fig. \ref{fig:fig10}(a). At 14 eV (blue symbols), this oscillation only extends over a few monolayers (typically 1 nm), which is understood from our previous discussion: at 14 eV, the DMI is dominated by the first nonmagnetic metal layer, resulting in oscillations confined close to the interface. In contrast, at 12.6 eV (red symbols), DMI does not saturate before about 20 monolayers, corresponding to about 2.3 nm, revealing the delocalized nature of DMI at this energy. Figure \ref{fig:fig10}(b) shows the dependence of DMI when varying the thickness of the ferromagnetic layer. At both Fermi energies, 14 eV and 12.6 eV, DMI saturates after a few monolayers only ($\sim 8$ monolayers, corresponding to less than 1 nm). This fast saturation is attributed to the spin dephasing, i.e., to the alignment of the spin of the delocalized electrons on the local magnetization of the ferromagnet. Due to the large exchange, any spin misalignment due to the magnetic texture is absorbed close to the interface, resulting in an interfacial behavior.

\section{Conclusion\label{s:conclusion}}
In this work, we discussed the nature of DMI in transition metal heterostructures. We first derived the expression of DMI in the weak spatial gradient limit within Keldysh formalism. This derivation provides us with a Green's function formula that is well adapted to tight-binding Hamiltonians. With this tool, we first uncover the role of orbital mixing and show that symmetry breaking enables the onset of interfacial orbital momentum that is at the origin of the DMI. We finally investigate the different layers to the DMI and reveal that it can expand over several nonmagnetic metal layers depending on the Fermi energy, thereby revealing the complex orbital texture of the band structure. Finally, we examine the thickness dependence of DMI on both ferromagnetic and nonmagnetic metal thicknesses and we find that whereas the former remains very weak, the thickness dependence of DMI as a function of the nonmagnetic metal thickness can be substantial.

\acknowledgments
This work was supported by the King Abdullah University of Science and Technology (KAUST) through the Office of
Sponsored Research (OSR) [Grant Number OSR-2017-CRG6-3390].  

\bibliography{Biblio2019}

\end{document}